\documentclass{article}
\usepackage{spconf,amsmath,graphicx}

\usepackage{cite}
\usepackage{amsmath,amssymb,amsfonts}
\usepackage{algorithmic}
\usepackage{graphicx}
\usepackage{textcomp}
\usepackage{xcolor}

\usepackage{bm}
\usepackage{multirow}
\usepackage{siunitx}

\usepackage[caption=false, font=footnotesize, position=bottom]{subfig}

\usepackage[firstpage=true]{background}
\SetBgContents{%
	\fbox{%
		\parbox{\textwidth}{%
			\footnotesize \textcopyright 2025 IEEE. Personal use of this material is permitted. Permission from IEEE must be obtained for all other uses, in any current or future media, including reprinting/republishing this material for advertising or promotional purposes, creating new collective works, for resale or redistribution to servers or lists, or reuse of any copyrighted component of this work in other works. }}%
}	
\SetBgScale{1}
\SetBgAngle{0}
\SetBgPosition{current page.north}
\SetBgVshift{-1.6cm}
\SetBgColor{black}
\SetBgOpacity{1}

\newcommand{\etal}{\textit{et~al.} }

\title{Variable Rate Learned Wavelet Video Coding Using Temporal Layer Adaptivity}
\name{Anna Meyer and Andr\'e Kaup \thanks{The authors gratefully acknowledge that this work has been funded by the Deutsche Forschungsgemeinschaft (DFG, German Research Foundation) under project number 461649014.}}
\address{
		\textit{Multimedia Communications and Signal Processing}\\
		\textit{Friedrich-Alexander-University Erlangen-Nürnberg}\\
		Erlangen, Germany}

\begin{document}
%
\maketitle
\begin{abstract}
Learned wavelet video coders provide an explainable framework by performing discrete wavelet transforms in temporal, horizontal, and vertical dimensions. With a temporal transform based on motion-compensated temporal filtering (MCTF), spatial and temporal scalability is obtained. In this paper, we introduce variable rate support and a mechanism for quality adaption to different temporal layers for a higher coding efficiency. Moreover, we propose a multi-stage training strategy that allows training with multiple temporal layers. Our experiments demonstrate Bj{\o}ntegaard Delta bitrate savings of at least -32\% compared to a learned MCTF model without these extensions. Training and inference code is available at: https://github.com/FAU-LMS/Learned-pMCTF. 

\end{abstract}
	\vspace{-2mm}
\section{Introduction}
\label{sec:intro}
	\vspace{-2mm}
The main paradigm in learned image and video compression is nonlinear transform coding \cite{Balle2021} similar to predictive transform coding in traditional compression.  Wavelet transforms have desirable properties for compression because their compromise between spatial and frequency resolution fits the correlation structure of visual data. Methods built on learned wavelet transforms also provide an explainable framework, as their latent space has a defined structure corresponding to a wavelet decomposition. In addition, they support lossless compression in contrast to most other learned methods. Learned wavelet coding is therefore specifically suited for tasks such as volumetric medical image coding \cite{Xue2021, Xue2023, Xue2024} and biomedical video coding \cite{Xue2023b}. 

In this paper, we address limitations of existing learned wavelet video coding schemes such as \cite{Meyer2023, Meyer2024a}.
These models use a trainable temporal wavelet transform realized using motion-compensated temporal filtering (MCTF). Learned MCTF supports spatial and temporal resolution scaling, where temporal scalability is achieved by a hierarchical group of pictures (GOP) structure with multiple temporal decomposition levels \cite{Ohm1994, Choi1999}. Up to now, learned MCTF coders require a separate model for every rate-distortion point \cite{Meyer2023, Meyer2024a}. This lack of support for continuous rate adaptation leads to increased training and storage costs. Moreover, existing learned MCTF models only support training on two frames. This results in a train-test mismatch, as during inference the models are evaluated with multiple temporal layers for GOPs larger than two. In addition, training on a single temporal layer does not allow quality adaptions to each layer, which gives considerable coding gains for traditional video coders in the context of B frame coding.
We tackle these issues with the following contributions: 
\begin{itemize}
		\vspace{-1mm}
	\item We enable smooth rate adjustment with trainable quality scaling parameters.
	\item We introduce temporal layer adaptive quality scaling considering the layer structure of learned MCTF for a higher coding efficiency.
	\item We propose a multi-stage training strategy that allows training on multiple temporal decomposition levels. This way, we increase the possible  training sequence length and eliminate the train-test mismatch for learned MCTF.
\end{itemize}
\begin{figure}[tb]
	\centering
	\includegraphics[width=0.45\textwidth]{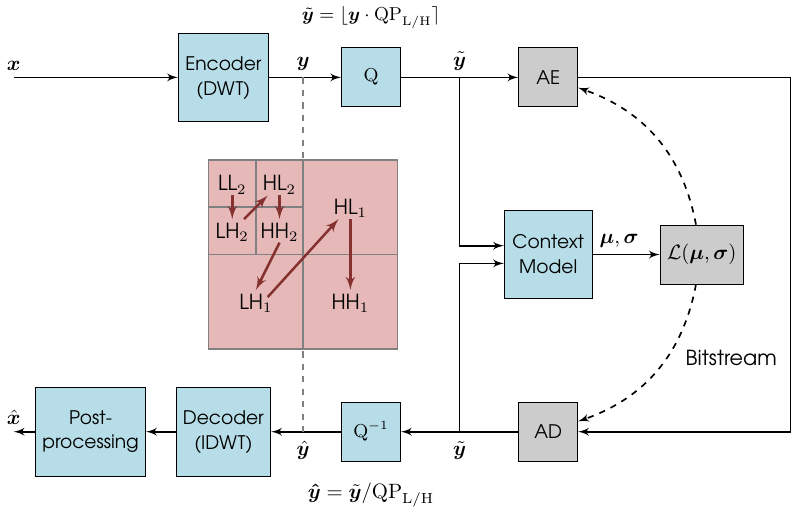}
		\vspace{-1mm}
	\caption{Overview of a learned wavelet image coder \cite{Meyer2024a}. $\bm{x}$ is one channel of an image in the YCbCr color space. The subbands $\bm{y}$ obtained from a learned discrete wavelet transform (DWT) are coded sequentially and their processing order is indicated by the red arrows.  }
	\label{fig:pwave}
	\vspace{-3.5mm}
\end{figure}
\begin{figure*}[tb]
	\subfloat[]{%
		\includegraphics[width=0.63\textwidth]{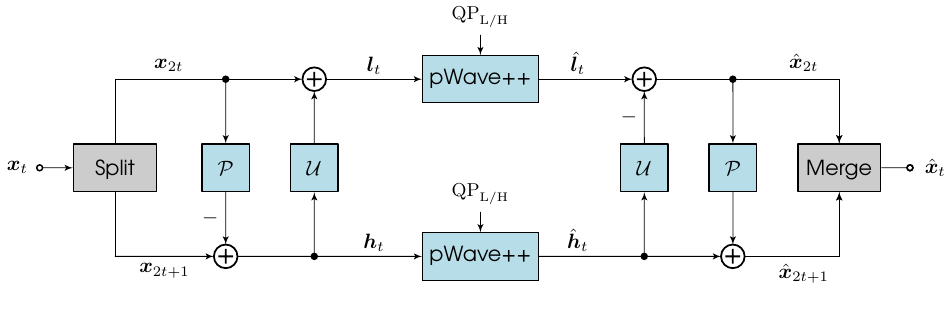} 	\label{fig:videomodel}
	}  {\color{lightgray}\vline}
	\subfloat[]{
		\includegraphics[width=0.35\textwidth]{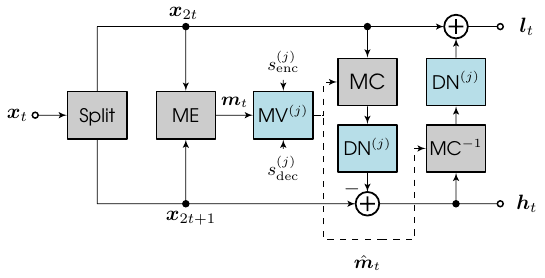}
		\label{fig:pandu}
	}
	\vspace{-2.5mm}
	\caption{ Overview of the wavelet video coder pMCTF. (a): Coding a single temporal level with two frames. (b): Implementation details of the predict and update filters. $\bm{x}_t$ denotes the input video. $\bm{h}_t$ and $\bm{l}_t$ are the temporal highpass and lowpass subband.  }
	\vspace{-2mm}
\end{figure*}
\vspace{-4.5mm}
\section{Related Work}
	\vspace{-1mm}
\subsection{Learned Wavelet Image Coding}
	\vspace{-1mm}
Ma \etal \cite{maiwave++} introduced a learned wavelet image coding framework called iWave++ that follows the structure visualized in Fig.~\ref{fig:pwave}. The encoding and decoding transforms are implemented using the lifting scheme. They perform a forward and inverse wavelet transform based on Convolutional Neural Networks (CNNs). A context model considering dependencies between subbands and within the subband to be coded allows iWave++ to achieve state-of-the-art performance. There have been several extensions of iWave++ \cite{Xue2023a, Xue2023} including a parallelized context model in pWave++ \cite{Meyer2024a}. Currently, there are ongoing efforts for standardizing wavelet image coders for the compression of regular images \cite{Dong2024} and medical image 3D volumes \cite{Xue2024}.
	\vspace{-3mm}
\subsection{Learned Wavelet Video Coding}
	\vspace{-2mm}
In the area of video coding, Dong \etal \cite{Dong2022} proposed a partly trainable framework with a traditional temporal wavelet transform and a subband coder following iWave++. In \cite{Meyer2023}, a fully trainable wavelet video coder based on MCTF was introduced. Its parallelized version pMCTF \cite{Meyer2024a} serves as the baseline model for our investigations on end-to-end wavelet video coding.
	
\textit{Model Overview:} Fig.~\ref{fig:videomodel} provides an overview of \mbox{pMCTF} \cite{Meyer2024a} for a single temporal decomposition level. A temporal wavelet transform realized via learned MCTF is followed by a spatial wavelet transform in pWave++, whereby all wavelet transforms are implemented using the lifting scheme. The lifting scheme consists of the three basic steps split, predict ($\mathcal{P}$) and update ($\mathcal{U}$). First, the input video sequence $\bm{x}_t$ is split into even and odd indexed frames $\bm{x}_{2t}$ and $\bm{x}_{2t+1}$. The temporal wavelet transform then computes a highpass and lowpass subband. The trainable predict step $\mathcal{P}$ performs a motion-compensated prediction to create the highpass subband containing residual information. The update step effectively performs lowpass filtering along the motion trajectory by reusing the computed motion vectors for inverse motion compensation on the highpass subband. The obtained temporal subbands are coded independently by dedicated pWave++ models with a parallelized four-step context model. The inverse temporal wavelet transform is obtained by reversing the lifting structure on the decoder side. 
\begin{figure}[tb]	
	\vspace{-3mm}
	\centering
	\includegraphics[width=0.25\textwidth]{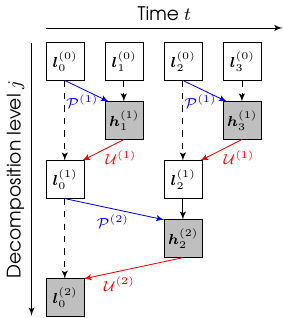}
	\caption{ Coding a GOP of size 4 using learned MCTF. The temporal lowpass and highpass subbands are denoted as $\bm{l}_{t}^{(j)}$ and $\bm{h}_{t}^{(j)}$. Encoding starts at the top with the first decomposition level $j=1$. After every level finished frames can be transmitted, as indicated by gray coloring.   }
	\label{fig:gop8}
	\vspace{-4mm}
\end{figure}

\textit{Hierarchical GOP Structure:} Learned MCTF relies on a dyadic decomposition that recursively applies a wavelet transform in temporal direction to the lowpass from the previous decomposition level. Hence, temporal scalability is provided with different temporal resolutions in different temporal layers. Fig.~\ref{fig:gop8} exemplarily shows the structure of a GOP of size 4 with 2 temporal layers. In the first layer, the predict filter $\mathcal{P}^{(1)}$ is applied to the original frames $\bm{l}_t^{(0)}$. The residual highpass frames $h_t^{(1)}$ and associated motion vectors can be directly transmitted. The update step  $\mathcal{U}^{(1)}$ gives the temporal lowpass frames $\bm{l}_t^{(1)}$  and the first level is finished. This temporal decomposition is repeated until there is only a single lowpass frame left. Every level has dedicated predict and update filters to account for different frame characteristics: $\mathcal{P}^{(1)}$ receives original frames as input, while one update step has been applied to the input of $\mathcal{P}^{(2)}$. 

\textit{Predict and Update Filters:} Fig.~\ref{fig:pandu} provides details on the implementation of the predict and update filters. For optical flow estimation, we use SpyNet \cite{Ranjan2017} to obtain the motion vectors $\bm{m}_t$.  For the motion vector compression module MV$^{(j)}$ we adopt the architecture from \cite{Li2023}. Based on the reconstructed motion vectors $\hat{\bm{m}}_t$, motion compensation (MC) is performed in the predict step and inverse motion compensation (MC$^{-1}$) in the update step. Both the predict and update filter contain a trainable denoising module (DN), which has the same residual CNN structure as the predict and update filters used in the spatial lifting structure of pWave++. All trainable blue components in Fig.~\ref{fig:pandu} depend on the temporal decomposition level $j$. Thus, there are dedicated modules for every temporal level.

However, pMCTF only supports training on two frames and different temporal levels are merely emulated by varying frame distances. This results in a train-test mismatch, because GOPs larger than two are used for testing and the hierarchical temporal layer structure cannot be considered during training. 
	\vspace{-5mm}
\section{Proposed Method}\vspace{-1mm}
\noindent
In the following, we introduce our proposed model which we refer to as pMCTF-L. Our model has three novel enhancements compared to the pMCTF model described above: Variable rate support,  temporal layer adaptive quality scaling, and a multi-stage training strategy enabling training with multiple layers.
	\vspace{-3mm}
\subsection{Variable Rate Support} \label{sec:quant}
	\vspace{-1mm} \noindent
Up to now, learned MCTF coders \cite{Meyer2023, Meyer2024a} require training dedicated models for every rate-distortion point. As popular approaches for continuous rate adaption like gain units \cite{cui2021asymmetric} are not compatible with the wavelet latent space, we adopt a different approach based on \cite{Li2024}. That way, we are able to introduce a unified rate adaption mechanism to both motion vector coding and the temporal subband coder pWave++. We let our proposed model pMCTF-L support different Quantization Parameters (QPs) in a predefined range with $q_{\mathrm{num}}$ = 21 values in total. During training, we randomly choose a discrete quantization index $q$ for every batch element. During inference, the quantization index can be selected from the continuous range [0, $q_{\mathrm{num}}-1$]. We interpolate the rate-distortion trade off parameter $\lambda$ according to \cite{Li2024}: 
\begin{equation}
	\lambda = \exp^{\ln \lambda_{\mathrm{min}} + \frac{q}{q_{\mathrm{num}} - 1} \cdot ( \ln \lambda_{\mathrm{max}} - \ln  \lambda_{\mathrm{min}} )}, \label{eq:int}
\end{equation}
where  $\lambda_{\mathrm{min}}$ and $ \lambda_{\mathrm{max}}$ define a lower and upper bound for $\lambda$. In our implementation, we choose $\lambda_{\mathrm{min}} = 0.03$ and $\lambda_{\mathrm{max}} = 0.081$ to cover a wide quality range.

The quantization index $q$ determines the quantization process inside the learned wavelet video coder as well: The motion vector encoder and decoder have dedicated scaling parameters $s_{\mathrm{enc}}^{(j)}$ and $s_{\mathrm{dec}}^{(j)}$ for every temporal level (cf. Fig.~\ref{fig:pandu}), while the pWave++ models have two quantization parameters QP$_{\mathrm{L}}$ for the LL subband and QP$_{\mathrm{H}}$ for the remaining subbands (cf. Fig.~\ref{fig:videomodel}). Both the layer-specific parameters for motion vector coding  and the QPs of pWave++ are interpolated according to (\ref{eq:int}). Each parameter is within the range of an associated trainable lower and upper bound.  
	\vspace{-2mm}
\subsection{Temporal Layer Adaptive Quality Scaling} 	\vspace{-1mm}
\noindent
Currently, learned MCTF coders do not consider their hierarchical temporal layer structure for bitrate allocation, even though this can give coding gains as known from traditional video coders. Therefore, we introduce an additional scaling parameter that depends on the decomposition level $j$ to allow adaptively scaling the quality depending on the temporal layer.
We obtain a layer-dependent quantization parameter QP$_{\mathrm{L}}$ for the lowpass subband and QP$_{\mathrm{H}}$ for the highpass subbands as:
\begin{equation}
	\mathrm{QP}_{\mathrm{L}/\mathrm{H}}^{(j)} = \mathrm{QP}_{\mathrm{L}/\mathrm{H}} \cdot q_{\mathrm{scale}}^{(j)} ,  \end{equation}
with $j\in \left\{ 1, 2, 3\right\}$ for a GOP of size 8 with 3 temporal layers. Analogously to the quantization parameters,  $q_{\mathrm{scale}}^{(j)}$ is in the range of a lower and upper bound defined by trainable parameters and interpolated based on the quantization index $q$ as specified in (\ref{eq:int}). In our experiments, we observe that $q_{\mathrm{scale}}^{(j)}$ increases the highpass quality with every decomposition level $j$. Hence, the highpass in the latest decomposition level is coded with the highest quality. This behavior is sensible because this decomposition level influences the reconstruction of all frames from previous temporal levels.
	\vspace{-2mm}
\subsection{Multi-Stage Training Strategy}  \label{sec:training}
	\vspace{-1mm}
\bgroup
\begin{table}[tb]
	\caption{Proposed training schedule for a GOP size of 16. $d_{\mathrm{max}}$ denotes the maximum frame distance between frames in a training sample and $T$ the number of frames. ''All'' trainable parts include the temporal wavelet transform with motion vector (MV) coding and the pWave++ models.   }
		\vspace{-2mm}
	\renewcommand{\arraystretch}{1.1}
	\begin{center}
		\resizebox{0.5\textwidth}{!}{
			\begin{tabular}{l|l|l|l|l|l|l}
				\multirow{2}{*}{Stage}	& \multirow{2}{*}{Trainable parts}  & \multirow{2}{*}{$T$} &  \multirow{2}{*}{$d_{\mathrm{max}}$}  & \multirow{2}{*}{Loss}  & \multirow{2}{*}{Learning rate}  & \multirow{2}{*}{Epochs}  \\[-0.1cm]
				&	&   & &  &  &  \\
				\hline
				1&	MV coding & 2 & 1 	 & $D_{\mathrm{ME}}$ & $\num{1e-04}$ & 1 \\	
				2 &	MV coding & 2 & 1 	 & $D_{\mathrm{ME}}+R_{\mathrm{MV}}$  & $\num{1e-04}$ & 3 \\	
				\hline
				3 & 	All  &  2 & 1 	 & $\mathcal{L}_{\mathrm{full}}$  & $\num{1e-05}$ & 5 \\	
				4 & 	MCTF  &  2 & 4 	 & $\mathcal{L}_{\mathrm{full}}$  & $\num{5e-05}$ & 2 \\	
				5 & 	All  & 2 & 4 	 &  $\mathcal{L}_{\mathrm{full}}$   & $\num{1e-05}$ & 5 \\	
				\hline
				6 & 	All  & 4 & 2 	 &  $\mathcal{L}_{\mathrm{full}}$   & $\num{1e-05}$ & 3 	\\
					\hline
				
			7 & 	All  & 8 & 1 	 &  $\mathcal{L}_{\mathrm{full}}$   & $\num{1e-05}$ & 5\\ 	
				8 & 	All  & 16 & 1 	 &  $\mathcal{L}_{\mathrm{full}}$   & $\num{1e-05}$ & 5\\
			\end{tabular}
			\label{tab:training}
		}
	\end{center}
		\vspace{-6mm}
\end{table}
\egroup 
\noindent
Up to now, learned MCTF coders were only trained on two frames, corresponding to a single temporal layer. As training with multiple layers is not feasible from scratch, 
we propose a multi-stage training strategy that extends the schedules from \cite{Meyer2023} and \cite{Sheng2023}. Table~\ref{tab:training} contains an overview of all six training stages. We initialize SpyNet with a pretrained model and keep the network fixed during all training stages, as we find this helps with overall training stability. 
Throughout all training stages, we randomly select a quantization index from the range $[0, q_{\mathrm{num}}-1]$ and use interpolated values for $\lambda$ as well as the quantization parameters described in Section~\ref{sec:quant}.

In the first two training stages (cf. Table~\ref{tab:training}), only the motion vector coding modules are trainable. First, the loss is the distortion  $D_{\mathrm{ME}}$ between the frame to be predicted $\bm{x}_{2t+1}$ and the prediction $\mathrm{MC}(\bm{x}_{2t})$ and also contains the rate $R_{\mathrm{MV}}$ required for motion vector coding in the second stage. In the third training stage,  all network components are trainable. The loss is the full rate-distortion loss for $T=2$ frames:
\begin{equation}
	\mathcal{L}_{\mathrm{full}} =
	\sum_{t=0}^{T-1}	R_{\mathrm{all}, t} + \lambda \cdot D_{\mathrm{MSE}}(\bm{x}_t, \hat{\bm{x}}_t), 
\end{equation}
where $t$ denotes the frame index and the distortion term corresponds to the Mean Squared Error (MSE) between the original frame $\bm{x}_t$ and the reconstructed frame $\hat{\bm{x}}_t$. $R_{\mathrm{all}, t}$ consists of the rate for coding the temporal subbands using a pWave++ model and $R_{\mathrm{MV}}$ if the frame at index $t$ is a highpass. 

In training stages 4 and 5, we select a random frame distance between one and $d_{\mathrm{max}}$ in every training step with equal probability. Depending on the frame distance, we now use different temporal wavelet transforms to model different temporal layers. For example, a frame distance of four corresponds to level $j=3$ with $\mathcal{P}_3$ and $\mathcal{U}_3$. The pWave++ models are kept fixed in stage 4 and are trainable again in stage 5. We tried omitting these training stages, however, we found that they are necessary as preparation for more temporal layers.

In stage 6, we use four frames per batch element to perform two full temporal decomposition levels. By that, we can train the quality scaling parameter $q_{\mathrm{scale}}^{(j)}$ that enables quality adaption to different temporal layers.

In training stages 7 and 8, we switch to a custom training set with a sequence length of 16 frames. This long-sequence data set allows to increase the training GOP size incrementally from 8 to 16. 
	\vspace{-2mm}
\section{Experiments and Results}
\label{sec:exp}
	\vspace{-2mm}
\subsection{Experimental Setup}\noindent

\textit{Training:} We use the public Vimeo90K \cite{xue2019video} as training data set for a training sequence length of up to 4. For longer training sequences, we use a 16-frame training set extracted from the Vimeo90k source clips for which we provide scripts for dataset creation. During training, we crop patches of size $128 \times 128$ from the luma channel of the respective training sample. We use AdamW as optimizer and a batch size of 8. We train pWave++  according to \cite{Meyer2024a} for initializing the pWave++ models in our video coder.  Different from \cite{Meyer2024a}, we incorporate variable rate support as described above.

\textit{Evaluation:} 	\noindent
We follow the test settings from \cite{Meyer2023, Li2022} and compare different coders on the first 96 frames of each test sequence. Our first test data set is the widely used  UVG \cite{Mercat2020} data set containing seven sequences of size 1920$\times$1080. To consider a different resolution, we test on three sequences (\textit{CityAlley}, \textit{FlowerFocus}, \textit{FlowerKids}) from the UVG 4K \cite{Mercat2020} data set with a resolution of 3840$\times$2160. 
\begin{figure}[tb] 
	\centering
	\subfloat[UVG]{%
		\vspace{-4mm}
		\includegraphics[width=0.35\textwidth]{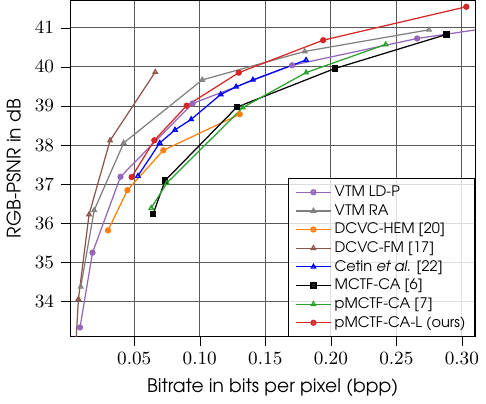}  \vspace{-0.3cm}
	}\\ \vspace{-0.3cm}
	\subfloat[UVG 4K]{		
		\includegraphics[width=0.3497\textwidth]{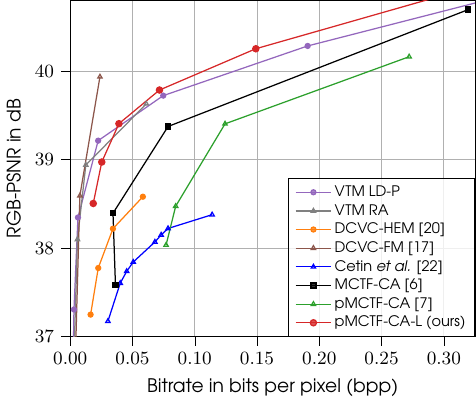}    \vspace{0.2cm}
		
	}	\vspace{-2mm}
	\caption{Rate-distortion evaluation with a GOP size of 16 on different data sets. }
	\vspace{-5mm} \label{fig:uvg4k_sota}
\end{figure}
	\vspace{-3mm}
\subsection{Comparison to State-of-the-Art Video Coders}
	\vspace{-2mm}
To evaluate state-of-the-art learned coders, we use DCVC-HEM \cite{Li2022} and DCVC-FM \cite{Li2024}. To consider a learned coding scheme different from conditional coding, we evaluate the bi-directional residual coding scheme by Cetin \etal \cite{Cetin2022}, which is an improved version of \cite{Yilmaz2022}. The coder follows a hierarchical resdidual coding structure like our MCTF model but supports bi-directional prediction. Because the GOP structure of MCTF is sensitive to the choice of GOP size, we allow a content adaptive GOP choice from \cite{Meyer2023} and obtain our model pMCTF-CA-L. In addition, we include other learned MCTF approaches in our experiments, namely MCTF-CA \cite{Meyer2023} and pMCTF-CA \cite{Meyer2024a}. The latter has a parallelized context model in comparison to MCTF-CA, but both models have no variable rate support and were trained with a single temporal layer only. We use GOP size of 16 for all models for a fair comparison to our approach.
To consider a traditional hybrid video coder, we include VTM~17.2 \cite{Bross2021}. We use VTM in Lowdelay P (LD-P) and Randomaccess (RA) configuration with an intra period of 16 such that it has the same I frame number as the learned coders.

The rate-distortion curves on the UVG and UVG 4K data set can be seen in Fig.~\ref{fig:uvg4k_sota}. The curves demonstrate that our \mbox{pMCTF-CA-L} model performs especially well for high rates where it outperforms VTM LD-P. Because of the model's capability of lossless compression it has the capacity to perform specifically well for higher rates. Also, our \mbox{pMCTF-CA-L} model clearly outperforms the other MCTF models which highlights the importance of adaptively considering the temporal layers structure of MCTF during training. Our model also performs better than DCVC-HEM and the residual bi-directional coder by Cetin \etal \cite{Cetin2022}. However, the conditional learned coder DCVC-FM shows the best performance compared to all learned as well as traditional coders. 

For a quantitative evaluation of the rate-distortion performance, we measure average Bj{\o}ntegaard Delta (BD) rate savings with VTM LD-P as an anchor in Table~\ref{tab:bd}. Using VTM as anchor allows a valid comparison as it covers the entire rate-distortion range of the other coders. We find that our pMCTF-CA-L model outperforms the other learned coders except for DCVC-FM in terms of PSNR as well as MS-SSIM on both data sets. Our model also outperforms VTM LD-P with unidirectional motion estimation on the UVG data set by achieving BD rate savings of -8\%. 
On the UVG 4K data set, the BD rates of the learned coders with VTM LD-P as an anchor can be much higher relative to the UVG data set. These large differences in performance show that it is sensible to test learned coders on high quality 4K content.
\begin{table}[tb]
	\vspace{-1mm}
	\centering
	\renewcommand{\arraystretch}{1.2}
	\caption{Rate-distortion evaluation on the UVG and UVG 4K data sets for a GOP size of 16. Average BD rate savings are provided relative to VTM LD-P as anchor.} 
	\vspace{0.7mm}	
	\resizebox{0.46\textwidth}{!}{%
		\begin{tabular}{l|c|c|c|c} 
			& \multicolumn{2}{c|}{RGB-PSNR} & \multicolumn{2}{c}{RGB-MS-SSIM}\\
			& UVG   	& UVG 4K 				& UVG   & UVG 4K	\\	\hline
			VTM RA 						&    -26.42\%   &   -28.6\%    	    &       -19.34\% &    -29.9\%    \\
			\hline
			DCVC-HEM \cite{Li2022} 		& 36.0\%	& 158.2\% 				& 35.4\% & 195.5\% \\
			DCVC-FM \cite{Li2024} 		& -43.7\%	&  -35.7\% 					& -36.1 \% & -40.5\% \\
			Cetin \etal \cite{Cetin2022} 	& 15.1\%	& 630\% 			&40.6\% & 541.6\% \\
			MCTF-CA \cite{Meyer2023}	& 51.0 \%  & 	192.30\%			&21.5\% & 104.7\%\\
			pMCTF-CA \cite{Meyer2024a}  & 56.9\%	 & 390.2\%   		&  31.3\%& 283.5\%\\
			pMCTF-CA-L (ours) 			& -8.0\%   & 12.0\%  					& -20.0\%& -33.0\% \\
		\end{tabular}
	}
	\vspace{-2mm}
	\label{tab:bd}
\end{table}
	\vspace{-3mm} 
	\begin{figure}[tb]
	\centering
	\includegraphics[width=0.35\textwidth]{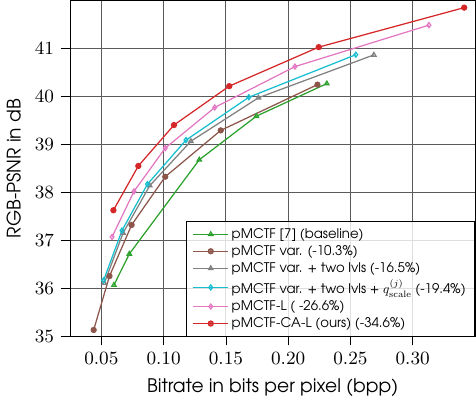} \vspace{-0.2cm}
	\vspace{-1mm}
	\caption{Ablation study on the UVG data set.  }
	\vspace{-5mm} 	\label{fig:ablation} 
\end{figure}
	\vspace{-2mm}
\subsection{Ablation Study}
	\vspace{-2mm}
We demonstrate the effectiveness of our approach by an ablation study on the UVG data set at a GOP size of 8. We take the pMCTF model from \cite{Meyer2024a} as baseline which was designed for a maximum GOP size of 8. Fig.~\ref{fig:ablation} provides the rate-distortion curves of the baseline pMCTF model and different configurations of our method. The brown~curve ''\mbox{pMCTF} var.'' achieves average BD rate savings of -10.3\% by enabling variable rate support, fixing the weights of the optical flow network, and using a four-step context model for motion vector coding. The model ''pMCTF var. + two lvls'' gives BD rate savings of -16.5\% by training on two temporal levels. With the temporal layer adaptive quality scaling, we get the model ''pMCTF var. + two lvls + $q_{\mathrm{scale}}^{(j)}$'' achieving BD rate savings of -19.4\%. When further increasing the training sequence length to 8 and enabling a latent prior for motion vector coding, we obtain our model pMCTF-L with BD rate savings of -26.6\%. The red curve corresponding to pMCTF-CA-L with layer adaptivity and a content adaptive GOP choice achieves average BD rate savings of -34.6\% over the baseline pMCTF model. 
	\vspace{-4mm}
\subsection{Complexity} 
	\vspace{-2mm}
\begin{table}[tb]
	\centering 
	\caption{ Complexity comparison in terms of kMAC/px and decoding time. Runtime is measured on a NVIDIA RTX A6000 GPU on center crops of size $512 \times 768$ from the UVG sequences. }
	\vspace{0.7mm}
	\renewcommand{\arraystretch}{1.2}
	\resizebox{0.38\textwidth}{!}{%
		\begin{tabular}{l|c|c}
			Model   & kMAC/px  & Decoding time\\
			\hline
			MCTF-CA \cite{Meyer2023}  & 3554 & 1865.6s  \\
			pMCTF-CA \cite{Meyer2024a}  &6593 & 7.7s\\
			pMCTF-CA-L (ours) &5401 & 8.7s\\
		\end{tabular}
	}
	\label{tab:time}	\vspace{-5mm}
\end{table}
In Table \ref{tab:time}, we assess the complexity of our method compared to other MCTF-based approaches. To this end, we measure average decoding times and complexity in terms of kilo multiply-accumulate operations per pixel (kMAC/px). The pMCTF models require more MACs due to the four-step context fusion models. By reducing the number of channels within this context model, our \mbox{pMCTF-L} model needs less MACs than pMCTF. Our model is more complex than the MCTF model from \cite{Meyer2023}, however, the average decoding time is significantly lower, which makes our pMCTF-L model most practical. 
	\vspace{-5mm}
\section{Conclusion} 
	\vspace{-3mm}
In this paper, we introduced variable rate support for learned MCTF and temporal layer adaptive quality scaling. By presenting a training strategy that supports multiple temporal layers we closed the train-test gap for learned MCTF coders. Overall, our proposed coder pMCTF-CA-L achieves BD rates savings of at least -32\% compared to pMCTF-CA for a GOP size of 16. This makes our coder competitive to VTM LD-P and advances learned wavelet video coders as an alternative learned video coding scheme with a known structure.
In future work, we will explore ways of considering information from multiple reference frames in the MCTF framework. 

	\vspace{-1mm}
	\ninept
\bibliographystyle{IEEEbib}
\bibliography{wavelets.bib}

\end{document}